\documentclass[prd,superscriptaddress,preprintnumbers,twocolumn,nofootinbib,showpacs]{revtex4-2}

\usepackage{amsfonts,amssymb,amsmath}
\usepackage{mathrsfs}
\usepackage{color}
\usepackage{graphicx}
\usepackage{dcolumn}
\usepackage{bm}
\usepackage{lipsum}
\usepackage{hyperref}

\begin{document}

\title{Probing inflationary moduli space with gravitational waves}
\author{Shinsuke Kawai}
\email{kawai@skku.edu}
\affiliation{
	Department of Physics, 
	Sungkyunkwan University,
	Suwon 16419, Republic of Korea}
\author{Jinsu Kim}
\email{kimjinsu@tongji.edu.cn}
\thanks{Corresponding Author}
\affiliation{
	School of Physics Science and Engineering, 
	Tongji University,
	Shanghai 200092, China}
\date{\today} 

\begin{abstract}
We investigate the spectrum of gravitational waves arising from primordial inflation in the presence of a string-theoretical higher curvature correction, specifically, the Gauss-Bonnet coupling term for the inflaton (modulus) field. We show that if the modulus field exhibits a wall-crossing like behavior in the moduli space, there can be a period of Gauss-Bonnet coupling term domination during the usual slow-roll. This phenomenon is potentially detectable as the gravitational wave spectrum exhibits a characteristic peak caused by the brief domination of the Gauss-Bonnet coupling term. We explore the possibility of measuring such gravitational waves with pulsar timing array experiments such as NANOGrav, and future space-borne interferometers such as LISA, DECIGO, and Taiji.
\end{abstract}

\maketitle

\section{Introduction}
\label{sec:intro}

The direct detection of gravitational waves by the Laser Interferometer Gravitational-Wave Observatory (LIGO) and Virgo Collaborations \cite{LIGOScientific:2016aoc,LIGOScientific:2016sjg,LIGOScientific:2017ycc} marked the beginning of gravitational wave astronomy.
The compelling signals of stochastic gravitational waves recently reported by the North American Nanohertz Observatory for Gravitational Waves (NANOGrav) \cite{NANOGrav:2023gor,NANOGrav:2023hde}, the Parkes Pulsar Timing Array (PPTA) \cite{Reardon:2023gzh,Zic:2023gta}, the European PTA (EPTA) \cite{EPTA:2023sfo,EPTA:2023fyk}, and the Chinese PTA (CPTA) \cite{Xu:2023wog} could be another breakthrough.
Although supermassive black hole binaries are the most likely source of these stochastic gravitational wave signals exhibiting the Hellings-Downs angular correlation, other cosmological sources have not been ruled out.
Various mechanisms have been proposed, including cosmic strings \cite{Ferrer:2023uwz,Ellis:2023tsl,Lazarides:2023ksx,Lazarides:2023rqf,Maji:2023fhv}, phase transitions \cite{Ashoorioon:2022raz,Fujikura:2023lkn,Bringmann:2023opz,Ghosh:2023aum}, scalar-induced gravitational waves (SIGWs) \cite{Franciolini:2023pbf,Das:2023nmm,Basilakos:2023xof,Cheung:2023ihl,HosseiniMansoori:2023mqh,Choudhury:2023wrm}, and inflation \cite{Vagnozzi:2023lwo,Jiang:2023gfe,Ben-Dayan:2023lwd,Choudhury:2023kam,Datta:2023vbs,Oikonomou:2023qfz,Borah:2023sbc}; see also Refs.~\cite{Madge:2023cak,Figueroa:2023zhu} for a detailed analysis on different cosmological sources.

One notable feature of gravitational wave astronomy is that through the direct observation of the gravitational sector, it enables us to test the theory of gravity.
In particular, gravitational waves of primordial origin are expected to contain the information about the physics beyond Einstein's gravity, as well as the state of the Universe before the Big Bang Nucleosynthesis, which is difficult to probe by other methods.
String-theoretical modifications of gravity involve scalar fields (moduli) associated with the extra dimensions, and it is natural to suppose that one of the moduli, that we call an inflaton, is responsible for cosmic inflation.
In this work, we will focus on a simple modification of Einstein's gravity with a scalar field coupled to the Euler density (often referred to as the Gauss-Bonnet term), which has been widely studied in the context of string theory \cite{Antoniadis:1992sa,Rizos:1993rt,Harvey:1995fq}; see also Refs.~\cite{Kawai:1997mf,Kawai:1999pw,Kawai:1998ab}.
Various aspects of the Gauss-Bonnet coupling term have been investigated in many different contexts; see, {\it e.g.}, Refs.~\cite{Nojiri:2005vv,Satoh:2007gn,Satoh:2008ck,Guo:2009uk,Guo:2010jr,Satoh:2010ep,Nojiri:2010wj,Koh:2014bka,Koh:2016abf,Nojiri:2017ncd,Tumurtushaa:2018agq,Koh:2018qcy,Odintsov:2018zhw,Lee:2020upe,Odintsov:2020mkz,Odintsov:2020xji,Odintsov:2020zkl,Chew:2020lkj,Zhang:2021rqs,Lee:2021uis,Oikonomou:2022xoq,Tsujikawa:2022aar,Kawaguchi:2022nku,Bayarsaikhan:2022rzc,Gangopadhyay:2022vgh,Khan:2022odn,Odintsov:2022cbm,Nojiri:2023jtf,Koh:2023zgn,Aoki:2023jvt,Biswas:2023eju}. 
Based on this framework, in Ref.~\cite{Kawai:2021edk}, we proposed a scenario for the formation of primordial black holes (PBHs) and determined the parameter space that can explain the observed dark matter abundance with PBHs. 
We also computed the spectrum of the SIGWs in this scenario and showed that they can be detected by future space-based interferometers, such as Laser Interferometer Space Antenna (LISA) \cite{LISA:2017pwj,Baker:2019nia}, DECi-hertz Interferometer Gravitational wave Observatory (DECIGO) \cite{Seto:2001qf,Kawamura:2006up}, Big Bang Observer (BBO) \cite{Crowder:2005nr,Corbin:2005ny,Harry:2006fi}, TianQin \cite{TianQin:2015yph}, and Taiji \cite{Ruan:2018tsw}, as well as ground-based observatories such as Square Kilometer Array (SKA) \cite{Carilli:2004nx,Janssen:2014dka,Weltman:2018zrl}.
In that model, large density fluctuations that lead to PBH formation and SIGW production are generated when the scalar field undergoes a transient ultra slow-roll phase near a fixed point that arises from the balance between the Gauss-Bonnet term and the potential term \cite{Kawai:2021bye}.

Below, we investigate the same model in different parameter regions, focusing on the spectrum of the gravitational waves that are generated.
We shall be interested in the inflaton dynamics which are opposite to the case of the effective ultra slow-roll \cite{Kawai:2021edk,Kawai:2021bye}, that is, the scalar field temporarily {\em accelerates} and {\em decelerates}, so the field velocity takes a large value for a brief period.
In this case, as we see in detail below, the transient large velocity period gives rise to a surge of gravitational wave amplitude, while the amplitude of the density perturbation remains small.
The model we study is motivated for example by a string-theoretical setup in which a modulus field exhibits a wall-crossing like behavior in the moduli space.
While the enhancement of gravitational waves in a similar setup was hinted at in Refs.~\cite{Satoh:2007gn,Guo:2009uk,Cai:2015dta,Cai:2015ipa,Cai:2015yza,Cai:2016ldn,Cai:2022nqv}, a detailed analysis on the gravitational wave spectrum is lacking to our knowledge, at least in this particular setup.
The aim of this paper is to compute the spectrum of the gravitational waves in this scenario and discuss the prospect of future detection.

The paper is organized as follows.
In Sec.~\ref{sec:model}, we present the model and discuss the background dynamics in the case where the scalar field temporarily accelerates. 
In Sec.~\ref{sec:primGWs}, we analyze the gravitational waves from inflation by solving the equation for the tensor perturbation both analytically and numerically. 
Section~\ref{sec:GWspec} presents today's gravitational wave spectrum predicted by this scenario and discusses the prospect for observation in the future. 
We conclude in Sec.~\ref{sec:conc} with brief comments. 
Technical details are summarized in two appendices.

\section{Model}
\label{sec:model}

We consider the action for the inflaton $\varphi$ and gravity,
\begin{align}
    S = \int d^4x \sqrt{-g} \left\{
    \frac{M_{\rm P}^2}{2}R
    -\frac{1}{2}(\partial \varphi)^2
    -V(\varphi)
    -\frac{\xi(\varphi)}{16}R_{\rm GB}^2
    \right\}\,,
\end{align}
where $M_{\rm P}$ is the reduced Planck mass, $R_{\rm GB}^2 \equiv R^2 - 4R_{\mu\nu}R^{\mu\nu} + R_{\mu\nu\rho\sigma}R^{\mu\nu\rho\sigma}$ is the four-dimensional Euler density (often referred to as the Gauss-Bonnet term), $V(\varphi)$ is the scalar potential, and $\xi(\varphi)$ is a coupling function which we shall choose
\begin{align}\label{eqn:tanh}
    \xi = \xi_0 \tanh[\xi_1(\varphi-\varphi_c)]
    \,,
\end{align}
where $\xi_0$, $\xi_1$ and $\varphi_c$ are real constant parameters.
This Gauss-Bonnet coupling is motivated by a wall-crossing like behavior of $\varphi$ in the moduli space \cite{Kawai:2021edk}.
In the context of string theory, the coupling of a modulus to the Euler density may arise for example as one-loop gravitational threshold corrections. 
For the type-II and heterotic examples that have been studied extensively \cite{Antoniadis:1992sa,Harvey:1995fq}, these are determined by the spectrum of Bogomol’nyi-Prasad-Sommerfield (BPS) states. 
When the spectrum suddenly changes,\footnote{
This is similar to a transition (such as supersymmetry breaking) in quantum field theory across which a beta function takes different values and the renormalization group flow changes. 
In our model, the transition (wall crossing) affects the equation of motion for the gravitational waves, leading to potentially observable consequences.
} that is, when the modulus $\varphi$ traverses a wall separating two domains with different BPS spectra, the coupling $\xi$ would behave as a step function at the wall location $\varphi_c$.
For a domain wall of a finite thickness, the function $\xi$ is modeled by a smeared step function, namely, Eq. \eqref{eqn:tanh} above.

The background equations are given by
\begin{align}
    0 &=
    \ddot{\varphi}
    +3H\dot{\varphi}
    +V_{,\varphi}
    +\frac{3}{2}H^2\left(H^2+\dot{H}\right)\xi_{,\varphi}
    \,,\label{eqn:KGeqn}\\
    0 &=
    3M_{\rm P}^2H^2
    -\frac{1}{2}\dot{\varphi}^2
    -V
    -\frac{3}{2}H^3\xi_{,\varphi}\dot{\varphi}
    \,,\label{eqn:Friedmanneqn}
\end{align}
where 
${}_{,\varphi}\equiv \partial/\partial\varphi$ and the dot represents the cosmic-time derivative.
In the early stage of inflation, when $\varphi$ is away from $\varphi_c$, the Gauss-Bonnet coupling term is negligible and the usual slow-roll inflation takes place.\footnote{
When $\xi$ is a constant, the Gauss-Bonnet term has no effects as it would be a topological term. 
Slow-roll inflation with Gauss-Bonnet corrections, with different forms of coupling function $\xi(\varphi)$, has been studied {\it e.g.} in Refs.~\cite{Satoh:2008ck,Guo:2010jr,Satoh:2010ep}.
} 
As the inflaton field approaches the critical value $\varphi_c$, the Gauss-Bonnet coupling term becomes significant as $\xi_{,\varphi}\sim {\rm sech}^2[\xi_1(\varphi-\varphi_c)]$.
If $V_{,\varphi}\xi_{,\varphi}<0$, a nontrivial fixed point may arise \cite{Kawai:2021bye} and abundant PBHs can be formed; also the enhanced SIGWs may become observable by near-future gravitational wave experiments \cite{Kawai:2021edk}.

We focus on the opposite, $V_{,\varphi}\xi_{,\varphi}>0$ case.
In this case, the Gauss-Bonnet coupling term can dominate over the potential term.
During this stage, {\it i.e.}, in the vicinity of $\varphi = \varphi_c$, the background inflaton equation of motion can be approximated as 
\begin{align}
    \dot{\varphi}^2 \simeq
    -\frac{1}{2}H^2M_{\rm P}^2(1-\epsilon_H)\sigma_1\,,
\end{align}
where $\epsilon_H \equiv -\dot{H}/H^2$ is the first Hubble slow-roll parameter and $\sigma_1 \equiv H\dot{\xi}/M_{\rm P}^2$ parametrizes the effects of the Gauss-Bonnet term. 
We thus see that the velocity gets enhanced when the inflaton crosses the critical value, as $\sigma_1$ becomes large.
From the time derivative of the Friedmann equation, we also obtain
\begin{align}
    \epsilon_H \simeq
    -\frac{\sigma_1\sigma_2}{4(1-\sigma_1/2)}
    \,,\label{eqn:e1s2}
\end{align}
where $\sigma_2\equiv \dot{\sigma}_1/(H\sigma_1)$.
Noting that $\sigma_2$ involves $\xi_{,\varphi\varphi} \sim -\tanh[\xi_1(\varphi-\varphi_c)]{\rm sech}^2[\xi_1(\varphi-\varphi_c)]$, we see that $\epsilon_H$ may become negative.
The brief domination of the Gauss-Bonnet coupling term is followed by the usual slow-roll inflation again.
As we shall discuss in the subsequent sections, the period of the Gauss-Bonnet domination may leave imprints on the gravitational wave spectrum.

\begin{figure*}
    \centering
    \includegraphics[scale=0.8]{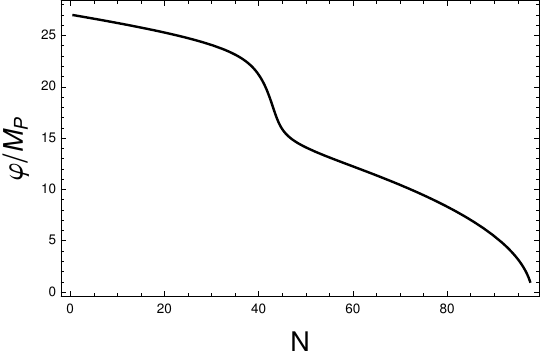}
    \includegraphics[scale=0.8]{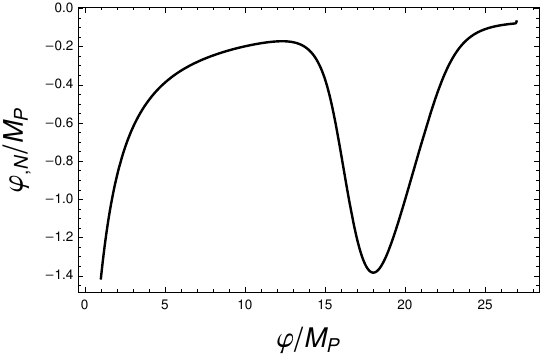}\\
    \includegraphics[scale=0.8]{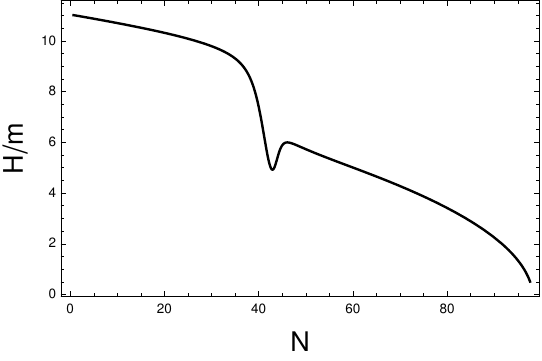}
    \includegraphics[scale=0.8]{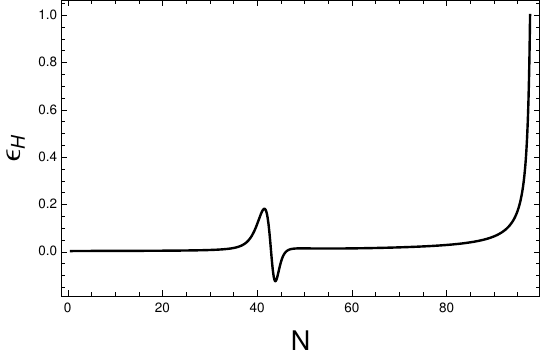}
    \caption{Evolution of the inflaton field (upper left), the inflaton velocity (upper right), the Hubble parameter (lower left), and the first Hubble slow-roll parameter $\epsilon_H = -\dot{H}/H^2$ (lower right); $N$ represents the number of $e$-folds. The parameters are chosen as $\{\xi_0,\xi_1,\varphi_c\}=\{0.196M_{\rm P}^2/m^2, 0.5/M_{\rm P}, 18.5 M_{\rm P}\}$ where $m$ is fixed by the amplitude of the curvature perturbation, $2.1\times 10^{-9}$, at the pivot scale which we take to be $N=70$ $e$-folds before the end of inflation. We observed that the Gauss-Bonnet coupling term dominates near $\varphi = \varphi_c$. Before and after the brief domination of the Gauss-Bonnet coupling term, the usual slow-roll inflation takes place.}
    \label{fig:bkg_solution}
\end{figure*}

For concreteness, we consider the potential of the form
\begin{align}\label{eqn:scalarpotential}
    V = \frac{1}{2}m^2\varphi^2\,.
\end{align}
Figure \ref{fig:bkg_solution} shows a prototypical behavior of the background solution. 
The upper-left panel represents the evolution of the scalar field in terms of the number of $e$-folds, whereas the upper-right panel shows the inflaton velocity. 
One can see that the inflaton temporarily accelerate and the velocity gets enhanced during the Gauss-Bonnet domination, {\it i.e.}, near $\varphi = \varphi_c$. In the lower-left (-right) panel, the behavior of the Hubble parameter (the first Hubble slow-roll parameter) is depicted. As $\epsilon_H$ temporarily becomes negative, the Hubble parameter shows an increasing behavior in the vicinity of the critical point $\varphi_c$. Before and after the brief domination of the Gauss-Bonnet term, the usual slow-roll inflation takes place. 
We use the quadratic form of the scalar potential \eqref{eqn:scalarpotential} for the sake of simplicity and the generic features described below do not depend on the details of the potential once the Gauss-Bonnet term dominates.

\section{Primordial Gravitational Waves}
\label{sec:primGWs}

Let us consider the tensor perturbation around the flat Friedmann-Lema\^itre-Robertson-Walker (FLRW) background metric in the exponential gauge,
\begin{align}
	ds^2 = 
	-dt^2
	+a^2[e^{h}]_{ij}dx^i dx^j
	\,.
\end{align}
We define
\begin{align}
	h_{ij} = h e_i e_j + \bar{h} \bar{e}_i \bar{e}_j\,,
\end{align}
where $e_i$ and $\bar{e}_i$ are the complex dyad that satisfy $\delta^{ij}e_i \bar{e}_j = 1$, $\delta^{ij}e_ie_j=\delta^{ij}\bar{e}_i\bar{e}_j=0$, $p^ie_i=p^i\bar{e}_i=0$, and $\epsilon^{ijk}p_ie_j\bar{e}_k=-ip$ for a given 3-momentum $p^i$. The Fourier transform of $h$, denoted by $h_{\bf k}$, satisfies \cite{Kawai:2017kqt}
\begin{align}
	h_{\bf k}^{\prime\prime}
	+2\frac{A_{t}^{\prime}}{A_{t}}
	h_{\bf k}^{\prime}
	+k^{2}C_{t}^{2}h_{\mathbf{k}} = 0\,,
\end{align}
where the prime denotes the conformal-time derivative, and
\begin{align}
	A_{t}^{2} &\equiv
	a^{2}\left(
    1 - \frac{\sigma_1}{2}
    \right)
	\,,\label{eqn:AtSquare}\\
	C_{t}^{2} &\equiv
	1 + \frac{a^2 \sigma_1}{2A_t^2}\left(
		1 - \sigma_2 - \epsilon_H
	\right)\,.\label{eqn:CtSquare}
\end{align}
Expanding $h_{\bf k}$ with the mode functions
\begin{align}
    h_{\bf k} &= 
    a_{\bf k} \mathfrak{h}_{\bf k} 
    + \bar{a}_{-{\bf k}}^\dagger \bar{\mathfrak{h}}^{*}_{-{\bf k}}
    \,,\\
    \bar{h}_{\bf k} &= 
    a_{-{\bf k}}^\dagger \mathfrak{h}^*_{-{\bf k}} 
    + \bar{a}_{{\bf k}} \bar{\mathfrak{h}}_{{\bf k}}
    \,,
\end{align}
the tensor perturbation is canonically quantized using the ladder operators satisfying 
$[a_{\bf k},a^\dagger_{-{\bf q}}] = [\bar{a}_{\bf k},\bar{a}^\dagger_{-{\bf q}}] = (2\pi)^3\delta^{(3)}({\bf k}+{\bf q})$. 
Introducing $v_{\bf k} \equiv (M_{\rm P}/2) A_t \mathfrak{h}_{\bf k}$ and $\bar{v}_{\bf k} \equiv (M_{\rm P}/2) A_t \bar{\mathfrak{h}}_{\bf k}$, the tensor perturbation equation becomes
\begin{align}\label{eqn:tensor_pert_eom}
    v_{\bf k}'' + \omega_{\bf k}^2 v_{\bf k} = 0\,,
\end{align}
and similarly for $\bar{v}$, where
\begin{align}
    \omega_{\bf k} \equiv 
    \sqrt{
    k^2C_t^2 
    - \frac{A_t''}{A_t}
    }\,.
\end{align}
We assume the Bunch-Davies vacuum and thus in the subhorizon limit, $v_{\bf k} \simeq e^{-iC_t k \tau}/\sqrt{2C_t k}$.

\begin{figure}
    \centering
    \includegraphics[scale=0.8]{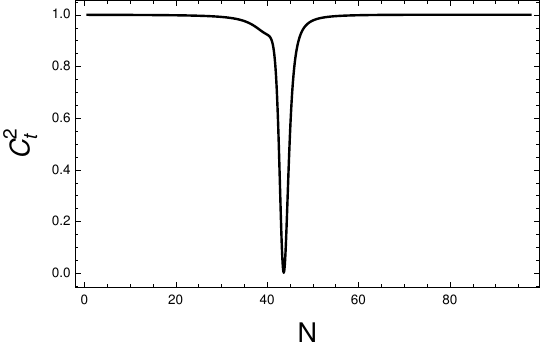}
    \caption{Behavior of $C_t^2$. The same parameter set is used as in Fig. \ref{fig:bkg_solution}. Near the critical point where the Gauss-Bonnet coupling term dominates, $C_t^2$ gets suppressed, while $C_t^2$ remains to be close to unity in the early and late stages of inflation where the usual slow-roll approximation holds.}
    \label{fig:CtSq}
\end{figure}

In the early and late stages of inflation, {\it i.e.}, away from the critical point, $A_t \simeq a$ and $C_t \simeq 1$, and we recover the standard tensor perturbation in the Einstein gravity. 
During the Gauss-Bonnet domination period, we find
\begin{align}
    C_t^2 &\simeq
    \frac{2}{2-\sigma_1}\left[
    1 + \left(2-\frac{3}{2}\sigma_1\right)\epsilon_H
    \right]
    \,.
\end{align}
Therefore, $C_t^2 \ll 1$ could be achieved\footnote{
For some parameter values $C_t$ can temporarily exceed unity.
This apparent superluminality of gravitational waves in the FLRW spacetime does not indicate violation of causality, as discussed, for example, in Refs.~\cite{deRham:2019ctd,deRham:2020zyh}.
} as shown in Fig. \ref{fig:CtSq}. It indicates that some modes may cross the horizon more than once.
In the vicinity of the critical point, we have
\begin{align}
    \frac{A_t''}{A_t} \simeq
    2a^2 H^2 \left[
    1 + \left(
    1 + \frac{\sigma_3}{2}
    \right)\epsilon_H
    \right]\,,
\end{align}
where $\sigma_3\equiv\dot{\sigma}_2/(H\sigma_2)$.
Using the conformal time (see Appendix \ref{apdx:conformaltime} for the derivation),
\begin{align}
    \tau &= \int \frac{da}{a^2H}
    \nonumber\\
    &= -\frac{1}{aH} - \frac{\epsilon_H}{aH}\left(
	\frac{\sigma_1}{4+\sigma_1}
	\right) + \mathcal{O}\left(\epsilon_H^2\right)
    \,,
	\label{eqn:conformaltime}
\end{align}
we see that the tensor perturbation equation \eqref{eqn:tensor_pert_eom} in the Gauss-Bonnet domination regime becomes
\begin{align}
    v_{\bf k}'' + \left[C_t^2k^2-\frac{\nu_t^2-1/4}{\tau^2}\right] v_{\bf k} = 0\,,
\end{align}
where
\begin{align}
    \nu_t \simeq \frac{3}{2} + \left[
    \frac{2(4+3\sigma_1)}{3(4+\sigma_1)}
    +\frac{\sigma_3}{3}
    \right]\epsilon_H\,.
\end{align}
The solution can be given by the Hankel function,
\begin{align}
	v_{\bf k}(\tau) =
	\frac{\sqrt{\pi}}{2}
	e^{i\frac{\pi}{2}\left(\nu_t+\frac{1}{2}\right)}
	\sqrt{-\tau}H_{\nu_t}^{(1)}\left(-C_tk\tau\right)\,.
\end{align}
Using the asymptotic form of the Hankel function in the superhorizon scale, we obtain the primordial tensor power spectrum as follows:
\begin{align}
    \mathcal{P}_{T,{\rm prim}}(k) &=
    \frac{4k^3}{\pi^2 M_{\rm P}^2A_t^2}\vert v_{\bf k} \vert^2
    =
	\mathcal{A}_T\left(
	\frac{C_tk}{aH}
	\right)^{3-2\nu_t}
    \,,
\end{align}
where the amplitude is given by
\begin{align}
	\mathcal{A}_T &=
	2^{2\nu_t}\left[
	\frac{\Gamma(\nu_t)}{\Gamma(3/2)}
	\right]^2
	\left(
	\frac{H}{2\pi M_{\rm P}}
	\right)^2
	\nonumber\\
	&\quad\times
	\frac{a^2}{A_t^2C_t^3}
	\left[
	1+\left(
	\frac{\sigma_1}{4+\sigma_1}
	\right)\epsilon_H
	\right]^{1-2\nu_t}\,.
\end{align}
In the regime of the Gauss-Bonnet domination, the tensor spectral index may thus become positive.
In addition, the amplitude of the tensor power spectrum peaks as the inflaton field approaches the critical point.
To the leading order, the amplitude can be approximated as 
\begin{align}
\mathcal{A}_T \approx \frac{H^2}{\pi^2M_{\rm P}^2}\sqrt{4+\frac{H^4}{M_{\rm P}^2}\xi_0^2\xi_1^2{\rm sech}^4[\xi_1(\varphi-\varphi_c)]}\,.
\end{align}
One may thus see that the amplitude is controlled by the product of the Gauss-Bonnet coupling parameters, $\mathcal{A}_T \sim \xi_0\xi_1$.

\begin{figure}
    \centering
    \includegraphics[scale=0.8]{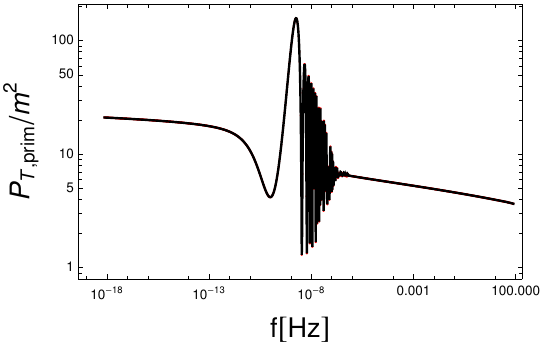}
    \caption{Primordial tensor power spectrum. The same parameter choice is used as in Fig. \ref{fig:bkg_solution}. The spectrum shows the enhancement and oscillatory behavior for modes that correspond to the Gauss-Bonnet domination period.}
    \label{fig:Pt_prim}
\end{figure}

We verify our findings by numerically solving the tensor perturbation equation \eqref{eqn:tensor_pert_eom}; see Appendix~\ref{apdx:perturbations} for details.
The result of the tensor power spectrum, normalized by $m^2$, for the parameter set used in the Fig.~\ref{fig:bkg_solution} is shown in Fig.~\ref{fig:Pt_prim}. For modes that correspond to the period of the Gauss-Bonnet domination, the primordial tensor power spectrum features a peak together with an oscillatory behavior as expected from the analysis discussed above.
We stress that the tensor perturbations in the current scenario differ from those arising in ultra slow-roll inflation scenarios. In the scenarios of ultra slow-roll inflation, sizable SIGWs accompany the enhancement of curvature perturbations. On the contrary, in the current scenario, no curvature perturbation enhancement occurs. Instead, the curvature power spectrum tends to be suppressed during the Gauss-Bonnet domination period, as shown in Fig.~\ref{fig:Pzeta_prim}; see Appendix~\ref{apdx:perturbations} for more details.
\begin{figure}
    \centering
    \includegraphics[scale=0.8]{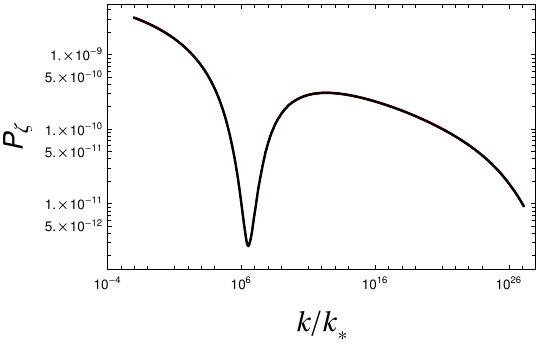}
    \caption{Primordial curvature power spectrum. The same parameter choice is used as in Fig. \ref{fig:Pt_prim}. The mass parameter $m$ is fixed to be $m\approx 3.65 \times 10^{-6} M_{\rm P}$ to match $\mathcal{P}_\zeta \approx 2.1\times 10^{-9}$ at the pivot scale $k_*=0.05 \, {\rm Mpc}^{-1}$. Unlike the SIGWs arising from ultra slow-roll inflation, no curvature perturbation enhancement occurs. Rather, it features suppression during the Gauss-Bonnet domination period.}
    \label{fig:Pzeta_prim}
\end{figure}

\section{Gravitational Waves Spectrum}
\label{sec:GWspec}

The present-day gravitational wave energy spectrum $\Omega_{\rm GW}$ is related to the primordial tensor power spectrum $\mathcal{P}_{T,{\rm prim}}$ through
\begin{align}\label{eqn:GWspectrum}
    \Omega_{\rm GW}(k) =
    \frac{1}{12}\left(
    \frac{k}{a_0 H_0}
    \right)^2
    T^2(k)
    \mathcal{P}_{T,{\rm prim}}
    \,,
\end{align}
where the ``0'' stands for the present-day values and the transfer function $T$ can be expressed as (see {\it e.g.}, Refs. \cite{Guzzetti:2016mkm, Kuroyanagi:2014nba,Kuroyanagi:2020sfw,Boyle:2005se})
\begin{align}
    T^2(k) &=
    \Omega_m^2
    \left(
    \frac{g_*(T_k)}{g_*^0}
    \right)
    \left(
    \frac{g_{*s}^0}{g_{*s}(T_k)}
    \right)^{4/3}
    \left(
    \frac{3j_1(k\tau_0)}{k\tau_0}
    \right)^2
    \nonumber\\
    &\quad\times
    \left[
    1+1.57\left(\frac{k_{\rm eq}}{k}\right)+3.42\left(\frac{k_{\rm eq}}{k}\right)^2
    \right]
    \nonumber\\
    &\quad\times
    \left[
    1-0.22\left(
    \frac{k_{\rm reh}}{k}
    \right)^{1.5}+0.65\left(
    \frac{k_{\rm reh}}{k}
    \right)^2
    \right]^{-1}
    \nonumber\\
    &\quad\times
    T_\nu^2
    \,.\label{eqn:transfer_function}
\end{align}
Here, $\Omega_m$ is the matter energy density, $j_1$ is the first spherical Bessel function, and $g_*$ ($g_{*s}$) is the effective relativistic (entropy) degrees of freedom. 
For the evolution of $g_*$ and $g_{*s}$, we assume the Standard Model of particle physics and use the following fitting function proposed in Ref. \cite{Kuroyanagi:2020sfw}:
\begin{align}
    g_*(T_k) &=
    g_*^0\left\{
    \frac{A+\tanh\left[
    -2.5\log_{10}\left(
    \frac{f}{2.5\times 10^{-12}\, {\rm Hz}}
    \right)
    \right]}{1+A}
    \right\}
    \nonumber\\
    &\quad\times
    \left\{
    \frac{B+\tanh\left[
    -2.0\log_{10}\left(
    \frac{f}{6.0\times 10^{-9}\, {\rm Hz}}
    \right)
    \right]}{1+B}
    \right\}
    \,,
\end{align}
and similarly for $g_{*s}$, where $f=k/(2\pi)$ is the frequency, and
\begin{gather}
    A = \frac{-1-10.75/g_*^0}{-1+10.75/g_*^0}
    \,,\quad
    B = \frac{-1-g_{\rm max}/10.75}{-1+g_{\rm max}/10.75}
    \,,\\
    g_*^0 = 3.36
    \,,\quad
    g_{*s}^0 = 3.91
    \,,\quad
    g_{\rm max} = 106.75
    \,.
\end{gather}
The horizon re-entry temperature is given by
\begin{align}
    T_k =
    5.8\times 10^6
    \left(
    \frac{g_{*s}(T_k)}{106.75}
    \right)^{-1/6}
    \left(
    \frac{k}{10^{14}\,{\rm Mpc}^{-1}}
    \right)\,{\rm GeV}
    \,,
\end{align}
and $k_{\rm eq}$ ($k_{\rm reh}$) is the wavenumber corresponding to the matter-radiation equality (end of reheating),
\begin{align}
    k_{\rm eq} &=
    7.1\times 10^{-2}
    \Omega_m h^2
    \,{\rm Mpc}^{-1}
    \,,\\
    k_{\rm reh} &=
    1.7\times 10^{14}
    \left(
    \frac{g_{*s}(T_{\rm reh})}{106.75}
    \right)^{1/6}
    \left(
    \frac{T_{\rm reh}}{10^7\,{\rm GeV}}
    \right)\,{\rm Mpc}^{-1}
    \,,
\end{align}
with $T_{\rm reh}$ being the reheating temperature; we assume the standard reheating scenario and take $T_{\rm reh} = 10^{14}$ GeV as an example.
Finally, $T_\nu$ takes into account the damping effect arising due to the free-streaming of neutrinos. We use the following fitting function \cite{Boyle:2005se}:
\begin{align}
    T_\nu &=
    \frac{15}{343(15+4f_v)(50+4f_v)(105+4f_v)(108+4f_v)}
    \nonumber\\
    &\quad\times
    \big(
    14406 f_v^4 - 55770 f_v^3 + 3152975 f_v^2 
    \nonumber\\
    &\quad\quad
    - 48118000 f_v^2 + 324135000
    \big)
    \,,
\end{align}
below the neutrino decoupling frequency, where $f_v=0.4052$ is the fraction of the neutrinos energy density.

\begin{figure}
    \centering
    \includegraphics[scale=0.8]{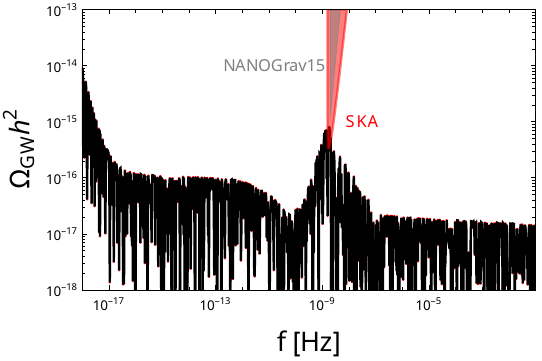}
    \caption{Gravitational wave energy spectrum obtained by using Eq.~\eqref{eqn:GWspectrum} for the primordial tensor power spectrum of Fig.~\ref{fig:Pt_prim}. The same parameter choice is used as in Fig.~\ref{fig:bkg_solution}. The spectrum peaks at nano-Hertz frequency range. The gray region depicts the latest observation of NANOGrav. The sensitivity curve for SKA is also depicted in red; the data is obtained from Ref.~\cite{Schmitz:2020syl}.}
    \label{fig:GWspectrum-nHz}
\end{figure}
Figure \ref{fig:GWspectrum-nHz} presents the gravitational wave energy spectrum for the parameter choice used in Fig. \ref{fig:Pt_prim}. We see that the spectrum peaks at the nano-Hertz range and thus may be probed by PTA experiments.

\begin{figure*}
    \centering
    \includegraphics[scale=0.8]{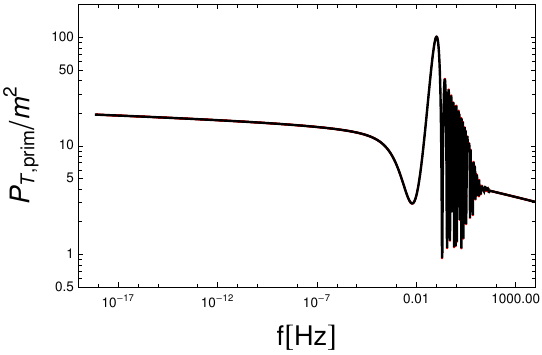}
    \includegraphics[scale=0.8]{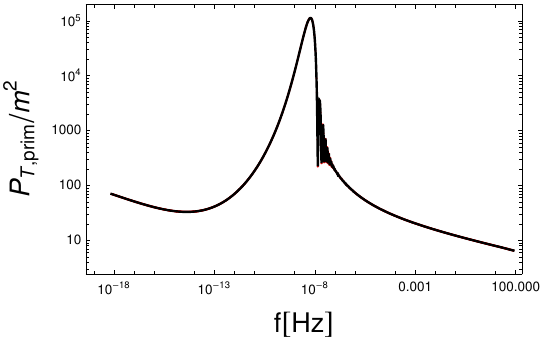}\\
    \includegraphics[scale=0.8]{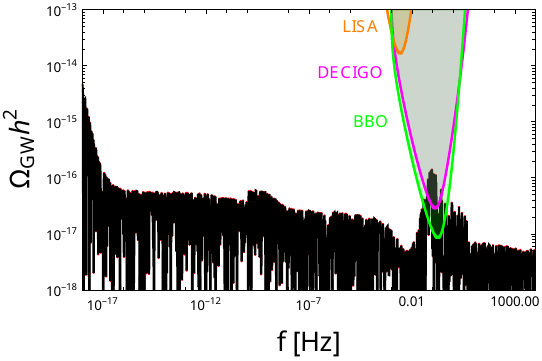}
    \includegraphics[scale=0.8]{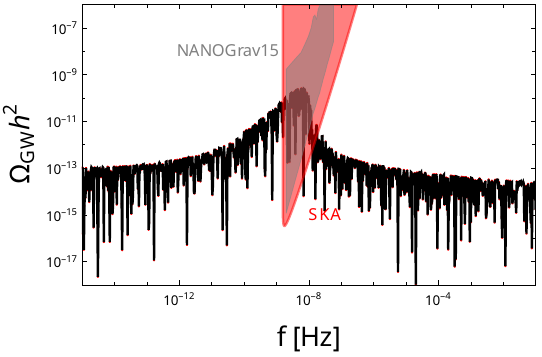}
    \caption{Primordial tensor power spectrum (upper panels) and gravitational wave energy spectrum (lower panels). The parameters are chosen as $\{\xi_0,\xi_1,\varphi_c\}=\{0.272M_{\rm P}^2/m^2, 0.5/M_{\rm P}, 15.5 M_{\rm P}\}$ (upper and lower left) and $\{\xi_0,\xi_1,\varphi_c\}=\{5.21M_{\rm P}^2/m^2, 0.02/M_{\rm P}, 73.0 M_{\rm P}\}$ (upper and lower right). The peak and oscillatory feature due to the Gauss-Bonnet domination appears in the Herz (nano-Herz) range for the former (latter) case. The gray region depicts the latest observation of NANOGrav. The sensitivity curves for SKA, LISA, DECIGO, and BBO are also depicted in red, orange, magenta, and green, respectively~\cite{Schmitz:2020syl}.}
    \label{fig:examples}
\end{figure*}

Before we conclude, we present two more examples in Fig.~\ref{fig:examples}. The primordial tensor power spectrum (present-day gravitational wave energy spectrum) is shown in the upper (lower) panel for two parameter sets, $\{\xi_0,\xi_1,\varphi_c\}=\{0.272M_{\rm P}^2/m^2, 0.5/M_{\rm P}, 15.5 M_{\rm P}\}$ (left panel) and $\{\xi_0,\xi_1,\varphi_c\}=\{5.21M_{\rm P}^2/m^2, 0.02/M_{\rm P}, 73.0 M_{\rm P}\}$ (right panel). The former case exhibits the Gauss-Bonnet domination feature in the Hertz frequency range, while the latter has the feature in the nano-Hertz frequency range. We observe that the enhancement is controlled by the combination of the Gauss-Bonnet coupling parameters, $\xi_0$ and $\xi_1$. The $\xi_0$ parameter, for a fixed value of $\xi_1$, regulates the suppression of $C_t^2$ and thus the oscillation period, while the $\xi_1$ parameter is responsible for the duration of the Gauss-Bonnet domination and thus the width of the peak. For given $\xi_0$ and $\xi_1$, the position of the wall, {\it i.e.}, $\varphi_c$, determines the peak position.

\section{Conclusion}
\label{sec:conc}

We have investigated primordial inflationary gravitational waves in the presence of the Gauss-Bonnet coupling function which typically arises in string-theoretical modification of gravity. Assuming the form of Gauss-Bonnet coupling function that models a wall-crossing like dynamics in the moduli space, we have shown that brief domination of the Gauss-Bonnet coupling term over the potential term is possible.

The primordial tensor fluctuations are enhanced by such temporary Gauss-Bonnet domination. We have presented analytical discussion for the enhancement and verified it by numerically solving the system of equations. Applying the transfer function representing the standard thermal history of the Universe after inflation to the primordial tensor power spectrum, we computed the present-day gravitational wave spectrum. The spectrum exhibits a characteristic peak.

Depending on the parameter values in the Gauss-Bonnet coupling function, the position of the peak in the gravitational wave spectrum varies in a wide frequency range. In this work, we have presented two cases; one where the peak is situated at the nano-Hertz frequency range and the other at the Hertz frequency range. The nano-Hertz case is thus in the prospective range of pulsar timing array experiments, such as NANOGrav, and especially of SKA, while the Hertz case may be probed by near-future gravitational wave observatories such as DECIGO and BBO.

Although the quadratic scalar potential was chosen in this work for the sake of simplicity and concreteness, we stress that the dynamics of the gravitational waves that we described in this work are insensitive to the choice of the potential, as long as the Gauss-Bonnet coupling term dominates the potential term such that $V_{,\varphi}\xi_{,\varphi} > 0$. Therefore, we suppose that the enhancement of the primordial inflationary gravitational waves may be realized for various other types of inflationary models once the Gauss-Bonnet coupling is included. Furthermore, we anticipate that a similar enhanced gravitational wave signal would arise in the presence of other higher curvature correction terms. We plan to investigate such scenarios in the future.

\begin{acknowledgments}
We acknowledge helpful communications with Sachiko Kuroyanagi and Jiro Soda.
S. K. is grateful for the hospitality of the theory group of the Helsinki Institute of Physics, University of Helsinki.
This work was supported in part by the National Research Foundation of Korea Grant-in-Aid for Scientific Research No. NRF-2022R1F1A1076172 (S. K.).
\end{acknowledgments}

\appendix
\section{CONFORMAL TIME}
\label{apdx:conformaltime}

In this appendix, we derive Eq. \eqref{eqn:conformaltime} for the conformal time $\tau = \int da/(a^2H)$.
Performing integration by parts, we find
\begin{align}
	\tau &= 
	-\frac{1}{aH} + \int \frac{\epsilon_H}{a^2H} da
	\,,
\end{align}
and
\begin{align}
	\int \frac{\epsilon_H}{a^2H} da &= 
	-\frac{\epsilon_H}{aH}
	+\int\frac{\epsilon_H\epsilon_2}{a^2H}da 
	+ \mathcal{O}\left(\epsilon_H^2\right)
	\,,
\end{align}
where $\epsilon_2 \equiv \dot{\epsilon}_H/(H\epsilon_H)$.
The time derivative of the Friedmann equation combined with the Klein-Gordon equation gives
\begin{align}
	\epsilon_H\epsilon_2 =
	\frac{\sigma_1}{4-\sigma_1}\left(
	\sigma_2-\sigma_2\sigma_3-2\epsilon_H
	\right)
	+\mathcal{O}\left(\epsilon_H^2\right)\,.
\end{align}
Thus,
\begin{align}
	\int\frac{\epsilon_H\epsilon_2}{a^2H}da &=
	-2\int\frac{\epsilon_H}{a^2H}\frac{\sigma_1}{4-\sigma_1}da
	\nonumber\\
	&\quad
	+\int\frac{da}{a^2H}\frac{\sigma_1\sigma_2(1-\sigma_3)}{4-\sigma_1}
	+\mathcal{O}\left(\epsilon_H^2\right)\,.
\end{align}
The first term can be integrated by parts to give
\begin{align}
	&\int\frac{\epsilon_H}{a^2H}\frac{\sigma_1}{4-\sigma_1}da
	\nonumber\\
	&=
	-\frac{\epsilon_H}{aH}\frac{\sigma_1}{4-\sigma_1}
	\nonumber\\
	&\quad
	+\int\frac{da}{a^2H}\left(
	\frac{\sigma_1}{4-\sigma_1}
	\right)^2(\sigma_2-\sigma_2\sigma_3-2\epsilon_H)
	\nonumber\\
	&\quad
	+\mathcal{O}\left(\epsilon_H^2\right)\,.
\end{align}
Thus,
\begin{align}
	&\int\frac{\epsilon_H\epsilon_2}{a^2H}da 
	\nonumber\\
	&=
	\frac{\epsilon_H}{aH}\frac{2\sigma_1}{4-\sigma_1}
	+4\int\frac{\epsilon_H}{a^2H}\left(
	\frac{\sigma_1}{4-\sigma_1}
	\right)^2da
	\nonumber\\
	&\quad
	+\int\frac{da}{a^2H}\left[
	\frac{\sigma_1\sigma_2}{4-\sigma_1}(1-\sigma_3)
	-2\left(
	\frac{\sigma_1}{4-\sigma_1}
	\right)^2
	\sigma_2(1-\sigma_3)
	\right]
	\nonumber\\
	&\quad
	+\mathcal{O}\left(\epsilon_H^2\right)
	\,.\label{eqn:eHe2int}
\end{align}
Integrating by parts the second term, we obtain
\begin{align}
	&\int\frac{\epsilon_H}{a^2H}\left(
	\frac{\sigma_1}{4-\sigma_1}
	\right)^2da
	\nonumber\\
	&=
	-\frac{\epsilon_H}{aH}\left(
	\frac{\sigma_1}{4-\sigma_1}
	\right)^2
	-2\int\frac{\epsilon_H}{a^2H}\left(
	\frac{\sigma_1}{4-\sigma_1}
	\right)^3da
	\nonumber\\
	&\quad
	+\int\frac{da}{a^2H}\frac{\sigma_1^3\sigma_2(1-\sigma_3)}{(4-\sigma_1)^3}
	\nonumber\\
	&\quad
	+\mathcal{O}\left(\epsilon_H^2\right)\,.
\end{align}
Substituting this result into Eq. \eqref{eqn:eHe2int}, we get
\begin{align}
	&\int\frac{\epsilon_H\epsilon_2}{a^2H}da 
	\nonumber\\
	&
	=
	\frac{\epsilon_H}{aH}\frac{2\sigma_1}{4-\sigma_1}
	-\frac{\epsilon_H}{aH}\left(
	\frac{2\sigma_1}{4-\sigma_1}
	\right)^2
	\nonumber\\
	&\quad
	-8\int\frac{\epsilon_H}{a^2H}\left(
	\frac{\sigma_1}{4-\sigma_1}
	\right)^3da
	\nonumber\\
	&\quad
	+\int\frac{da}{a^2H}\left[
	\frac{\sigma_1\sigma_2(1-\sigma_3)}{4-\sigma_1}
	\right]
	-2\int\frac{da}{a^2H}\left[
	\frac{\sigma_1^2\sigma_2(1-\sigma_3)}{(4-\sigma_1)^2}
	\right]
	\nonumber\\
	&\quad
	+4\int\frac{da}{a^2H}
	\frac{\sigma_1^3\sigma_2(1-\sigma_3)}{(4-\sigma_1)^3}
	+\mathcal{O}\left(\epsilon_H^2\right)\,.
\end{align}
The third term can again be integrated by parts. Repeating this process, we find
\begin{align}
	\int\frac{\epsilon_H\epsilon_2}{a^2H}da
	&=
	-\frac{\epsilon_H}{aH}s\left(
	1+s+s^2+\cdots
	\right)
	\nonumber\\
	&\quad
	-\int\frac{da}{a^2H}\frac{\sigma_2(1-\sigma_3)}{2}s\left(
	1+s+s^2+\cdots
	\right)
	\nonumber\\
	&\quad
	+\mathcal{O}\left(\epsilon_H^2\right)\,,
\end{align}
where $s \equiv 2\sigma_1/(\sigma_1-4)$.
Let us first consider the case where $s<1$, {\it i.e.}, $-4<\sigma_1<4/3$; note that $\sigma_1<0$ in our case.
In this case,
\begin{align}
	\int\frac{\epsilon_H\epsilon_2}{a^2H}da
	&=
	\frac{\epsilon_H}{aH}\left(
	\frac{2\sigma_1}{4+\sigma_1}
	\right)
	\nonumber\\
	&\quad
	+\int\frac{da}{a^2H}\frac{\sigma_1\sigma_2}{4+\sigma_1}
	-\int\frac{da}{a^2H}\frac{\sigma_1\sigma_2\sigma_3}{4+\sigma_1}
	\nonumber\\
	&\quad
	+\mathcal{O}\left(\epsilon_H^2\right)
	\nonumber\\
	&=
	\frac{\epsilon_H}{aH}\left(
	\frac{2\sigma_1}{4+\sigma_1}
	\right)
	-\frac{1}{aH}\frac{\sigma_1\sigma_2}{4+\sigma_1}
	\nonumber\\
	&\quad
	+\mathcal{O}\left(\epsilon_H^2\right)
	\,,
\end{align}
where we have performed integration by parts.
We thus conclude that
\begin{align}
	\tau =
	-\frac{1}{aH} - \frac{\epsilon_H}{aH}\left(
	\frac{\sigma_1}{4+\sigma_1}
	\right) + \mathcal{O}\left(\epsilon_H^2\right)
	\,.\label{eqn:conformaltimeapdx1}
\end{align}
where we have used Eq. \eqref{eqn:e1s2}.

Let us now consider the $s>1$ case. We take the limit $|\sigma_1| \gg 1$. In this case, since
\begin{align}
	\int\frac{\epsilon_H\epsilon_2}{a^2H} \simeq
	2\int\frac{\epsilon_H}{a^2H}da
	-\int\frac{da}{a^2H}\sigma_2(1-\sigma_3)\,,
\end{align}
up to the leading order, we find
\begin{align}
	\int\frac{\epsilon_H}{a^2H}da \simeq
	\frac{\epsilon_H}{aH}
	+\int\frac{\sigma_2}{a^2H}da
	-\int\frac{\sigma_2\sigma_3}{a^2H}da\,.
\end{align}
Integrating by parts the second term, we obtain
\begin{align}
	\tau \simeq
	-\frac{1}{aH}
	+\frac{\epsilon_H}{aH}
	-\frac{\sigma_2}{aH}
	\simeq
	-\frac{1}{aH}
	-\frac{\epsilon_H}{aH}
	\,,\label{eqn:conformaltimeapdx2}
\end{align}
where we have used Eq. \eqref{eqn:e1s2} in the $|\sigma_1|\gg1$ limit.
We note that Eq. \eqref{eqn:conformaltimeapdx1} captures Eq. \eqref{eqn:conformaltimeapdx2} in the $|\sigma_1|\gg1$ limit. We thus report Eq. \eqref{eqn:conformaltimeapdx1} in the main text.

\section{CURVATURE AND TENSOR PERTURBATIONS}
\label{apdx:perturbations}

Following Ref.~\cite{Kawai:2021bye} (see also Ref.~\cite{Kawai:2021edk}), we introduce the dimensionless quantities as
\begin{align}
    \bar{\varphi} \equiv \frac{\varphi}{M_{\rm P}}
    \,,\quad
    \bar{H} \equiv \frac{H}{m}
    \,,\quad
    \bar{V} \equiv \frac{V}{m^2M_{\rm P}^2}
    \,,\quad
    \bar{\xi} \equiv \frac{m^2}{M_{\rm P}^2}\xi
    \,.
\end{align}
We note that $\bar{\xi}=\bar{\xi}_0\tanh[\bar{\xi}_1(\bar{\varphi}-\bar{\varphi}_c)]$ where $\bar{\xi}_0 = (m^2/M_{\rm P}^2)\xi_0$ and $\bar{\xi}_1 = M_{\rm P}\xi_1$. 
Note that the Gauss-Bonnet coupling function $\xi$, and thus $\xi_0$ as well, are already dimensionless, but we have introduced $\bar{\xi}$ and $\bar{\xi}_0$ so that the mass parameter $m$ disappears as we shortly show.
For the time variable, we use the number of $e$-folds $N = \int H dt$ which is a dimensionless quantity.
In terms of the dimensionless variables, the background equations \eqref{eqn:KGeqn} and \eqref{eqn:Friedmanneqn} become
\begin{align}
    0 &=
    \bar{\varphi}_{,NN} + \left(
    3 + \frac{\bar{H}_{,N}}{\bar{H}}
    \right)\bar{\varphi}_{,N} + \frac{\bar{V}_{,\bar{\varphi}}}{\bar{H}^2} 
    \nonumber\\
    &\qquad
    + \frac{3}{2}\bar{ H}^2\left(
    1 + \frac{\bar{H}_{,N}}{\bar{H}}
    \right)\bar{\xi}_{,\bar{\varphi}}
    \,,\label{eqn:dimlessKGeqn}\\
    0 &=
    3 - \frac{\bar{V}}{\bar{H}^2} - \frac{1}{2}\bar{\varphi}_{,N}^2 - \frac{3}{2}\bar{H}^2\bar{\xi}_{,\bar{\varphi}}\bar{\varphi}_{,N}
    \,.\label{eqn:dimlessFriedmanneqn}
\end{align}
As advertised, the mass parameter $m$ drops off from the background equations.

The curvature perturbation equation is given by \cite{Kawai:2021bye,Kawai:2021edk}
\begin{align}
    u_{\bf k}'' + \left(
    C_\zeta^2 k^2 - \frac{A_\zeta''}{A_\zeta}
    \right)u_{\bf k} = 0
    \,,
\end{align}
where $u_{\bf k} = M_{\rm P} A_\zeta \zeta_{\bf k}$, $\zeta_{\bf k}$ is the Fourier transform of the curvature perturbation $\zeta$, and
\begin{align}
    A_\zeta^2 &=
    a^2\left(
    \frac{1-\sigma_1/2}{1-3\sigma_1/4}
    \right)^2
    \\
    &\qquad
    \times\left(
    2\epsilon_H - \frac{1}{2}\sigma_1
    + \frac{1}{2}\sigma_1\sigma_2
    - \frac{1}{2}\sigma_1\epsilon_H
    + \frac{3}{4}\frac{\sigma_1^2}{2-\sigma_1}
    \right)
    \,,\nonumber\\
    C_\zeta^2 &=
    1 - \frac{a^2}{A_\zeta^2}\left(
    \frac{\sigma_1}{2-3\sigma_1/2}
    \right)^2
    \\
    &\qquad
    \times\left(
    2\epsilon_H + \frac{1}{4}\sigma_1
    - \frac{1}{4}\sigma_1\sigma_2
    - \frac{5}{4}\sigma_1\epsilon_H
    \right)
    \,.\nonumber
\end{align}
In terms of the dimensionless variables, the curvature perturbation equation becomes
\begin{align}
    0 &=
    \bar{u}_{{\bf k},NN}
    + \left(
    1 + \frac{\bar{H}_{,N}}{\bar{H}}
    \right)\bar{u}_{{\bf k},N}
    + \frac{1}{a^2\bar{H}^2}\bigg\{
    C_\zeta^2 \bar{k}^2
    \nonumber\\
    &\qquad
    -\bigg[
    2 
    + \frac{3}{2}\frac{\tilde{A}_{\zeta,N}}{\tilde{A}_\zeta}
    - \frac{1}{4}\left(\frac{\tilde{A}_{\zeta,N}}{\tilde{A}_\zeta}\right)^2
    + \frac{1}{2}\frac{\tilde{A}_{\zeta,NN}}{\tilde{A}_\zeta}
    \nonumber\\
    &\qquad
    + \frac{\bar{H}_{,N}}{\bar{H}}
    + \frac{1}{2}\frac{\tilde{A}_{\zeta,N}\bar{H}_{,N}}{\tilde{A}_\zeta\bar{H}}
    \bigg]
    \bigg\}\bar{u}_{\bf k}
    \,,\label{eqn:dimlessCurvPerteqn}
\end{align}
where we have defined
\begin{align}
    \tilde{A}_\zeta \equiv \frac{A_\zeta^2}{a^2}
    \,,\quad
    \bar{k} \equiv \frac{k}{m}
    \,,\quad
    \bar{u}_{\bf k} \equiv \sqrt{m} u_{\bf k}
    \,.
\end{align}
Note that $A_\zeta$, and thus $\tilde{A}_\zeta$, as well as $C_\zeta$ are already dimensionless. Note also that the mass parameter $m$ does not enter the equation.
We solve the dimensionless curvature perturbation equation \eqref{eqn:dimlessCurvPerteqn} together with the dimensionless background equations \eqref{eqn:dimlessKGeqn} and \eqref{eqn:dimlessFriedmanneqn} with the standard Bunch-Davies vacuum state in the far subhorizon limit,
\begin{align}
    \lim_{\tau \rightarrow -\infty} \bar{u}_{\bf k}(\tau) = \frac{1}{\sqrt{2C_\zeta \bar{k}}}e^{-i C_\zeta k \tau}
    \,,
\end{align}
as the initial condition.
The curvature power spectrum is then given by
\begin{align}
    \mathcal{P}_\zeta = \frac{k^3}{2\pi^2}|\zeta_{\bf k}|^2
    = \frac{m^2\bar{k}^3}{2\pi^2 a^2}\frac{|\bar{u}_{\bf k}|^2}{M_{\rm P}^2\tilde{A}_\zeta}
    \,,
\end{align}
evaluated in the far superhorizon limit. The mass parameter $m$ is then found by imposing the normalization condition that $\mathcal{P}_\zeta \approx 2.1 \times 10^{-9}$ at the pivot scale $k_* = 0.05\,{\rm Mpc}^{-1}$.

The tensor power spectrum can be computed in the same manner.
The tensor perturbation equation is given by 
\begin{align}
    0 &=
    \bar{v}_{{\bf k},NN}
    + \left(
    1 + \frac{\bar{H}_{,N}}{\bar{H}}
    \right)\bar{v}_{{\bf k},N}
    + \frac{1}{a^2\bar{H}^2}\bigg\{
    C_t^2 \bar{k}^2
    \nonumber\\
    &\qquad
    -\bigg[
    2 
    + \frac{3}{2}\frac{\tilde{A}_{t,N}}{\tilde{A}_t}
    - \frac{1}{4}\left(\frac{\tilde{A}_{t,N}}{\tilde{A}_t}\right)^2
    + \frac{1}{2}\frac{\tilde{A}_{t,NN}}{\tilde{A}_t}
    \nonumber\\
    &\qquad
    + \frac{\bar{H}_{,N}}{\bar{H}}
    + \frac{1}{2}\frac{\tilde{A}_{t,N}\bar{H}_{,N}}{\tilde{A}_t\bar{H}}
    \bigg]
    \bigg\}\bar{v}_{\bf k}
    \,,\label{eqn:dimlessTenPerteqn}
\end{align}
where
\begin{align}
    \tilde{A}_t \equiv \frac{A_t^2}{a^2}
    \,,\quad
    \bar{k} \equiv \frac{k}{m}
    \,,\quad
    \bar{v}_{\bf k} \equiv \sqrt{m} v_{\bf k}
    \,.
\end{align}
Here, $\bar{v}_{\bf k}$, which denotes the dimensionless version of $v_{\bf k}$, is not to be confused with the one defined in Sec.~\ref{sec:primGWs}.
Note that $A_t$ and $C_t$, whose expressions are given by Eqs.~\eqref{eqn:AtSquare} and \eqref{eqn:CtSquare}, are already dimensionless. Also, the mass parameter $m$ does not enter the equation.
We again solve the dimensionless tensor perturbation equation \eqref{eqn:dimlessTenPerteqn} together with the dimensionless background equations \eqref{eqn:dimlessKGeqn} and \eqref{eqn:dimlessFriedmanneqn} with the standard Bunch-Davies vacuum state in the far subhorizon limit,
\begin{align}
    \lim_{\tau \rightarrow -\infty} \bar{v}_{\bf k}(\tau) = \frac{1}{\sqrt{2C_t \bar{k}}}e^{-i C_t k \tau}
    \,,
\end{align}
as the initial condition.
The tensor power spectrum is
\begin{align}
    \mathcal{P}_{T,{\rm prim}} = 2\times \frac{k^3}{2\pi^2}|h_{\bf k}|^2
    = \frac{4m^2\bar{k}^3}{\pi^2 a^2}\frac{|\bar{v}_{\bf k}|^2}{M_{\rm P}^2\tilde{A}_t}
    \,,
\end{align}
evaluated in the far superhorizon limit.


\end{document}